\documentclass[twocolumn,superscriptaddress,amsmath,amssymb,aps,prl,floatfix,nobibnotes]{revtex4-1}

\usepackage{graphicx}
\usepackage{dcolumn}
\usepackage{bm}

\usepackage{xcolor}
\usepackage{comment}

\begin{document}

\title{Magneto-Optical Measurements of the Negatively Charged 2$s$ Exciton in WSe$_2$}

\author{J.C. Sell*}
\affiliation{Joint Quantum Institute, NIST and University of Maryland, College Park, MD, 20742, USA}
\thanks{These two authors contributed equally}
\author{J.R. Vannucci*}
\affiliation{Joint Quantum Institute, NIST and University of Maryland, College Park, MD, 20742, USA}
\thanks{These two authors contributed equally}
\author{D.G. Su\'{a}rez-Forero}
\affiliation{Joint Quantum Institute, NIST and University of Maryland, College Park, MD, 20742, USA}
\author{B. Cao}
\affiliation{Joint Quantum Institute, NIST and University of Maryland, College Park, MD, 20742, USA}
\author{D.W. Session}
\affiliation{Joint Quantum Institute, NIST and University of Maryland, College Park, MD, 20742, USA}
\author{H.-J. Chuang}
\affiliation{Nova Research, Inc., Washington, DC 20375, USA}
\thanks{At the time of submission, H.-J.C. has transitioned to a position at the Naval Research Laboratory}
\author{K.M. McCreary}
\affiliation{Naval Research Laboratory, Washington, DC 20375, USA}
\author{M.R. Rosenberger}
\affiliation{Department of Aerospace and Mechanical Engineering, University of Notre Dame, Notre Dame, IN 46556, USA}
\author{B.T. Jonker}
\affiliation{Naval Research Laboratory, Washington, DC 20375, USA}
\author{S. Mittal}
\affiliation{Department of Electrical and Computer Engineering, Northeastern University, Boston, MA 02115, USA}
\author{M. Hafezi}
\affiliation{Joint Quantum Institute, NIST and University of Maryland, College Park, MD, 20742, USA}
\affiliation{Department of Electrical Engineering and Institute for Research in Electronics and Applied Physics, University of Maryland, College Park, MD 20742, USA}

\begin{abstract}
Monolayer transition metal dichalcogenides  (TMDs) host a variety of optically excited quasiparticle species that stem from two-dimensional confinement combined with relatively large carrier effective masses and reduced dielectric screening. The magnetic response of these quasiparticles gives information on their spin and valley configurations, nuanced carrier interactions, and insight into the underlying band structure. Recently, there have been several reports of 2$s$/3$s$ charged excitons in TMDs, but very little is still known about their response to external magnetic fields. Using photoluminescence excitation spectroscopy,  we observe the presence of the 2$s$ charged exciton and report its response to an applied magnetic field. We benchmark this response against the neutral exciton and find that both the 2$s$ neutral and charged excitons exhibit similar behavior with $g$-factors of  g$_{\rm{X_0^{2s}}}$=-5.20$\pm0.11$ and g$_{\rm{X_-^{2s}}}$=-4.98$\pm0.11$, respectively.
\end{abstract}

\maketitle

Monolayer semiconductor transition metal dichalcogenides (TMDs) have attracted significant attention in the last decade due to their unique optical properties. Similar to graphene, but with a three-layer (staggered) honeycomb lattice, TMDs host direct-gap transitions at their $\pm$K valleys and exhibit circular-dichroism due to their finite Berry curvature \cite{Mak-KF2,Xiao-D,Mak-KF3}. The reduced dimensionality of materials in this system, coupled with techniques like hexagonal boron nitride (hBN) encapsulation, lead to enhanced Coulomb interaction and excitons with large binding energies ($E_{\rm{B}}  \approx 150-500$ $\rm{ meV})$ \cite{Stier-AV,Chernikov-A2,He-K}. 

When there is excess charge present in the system during exciton formation, the exciton may lower its energy by capturing an electron or hole and form a bound, charged three-body state referred to as a charged exciton \cite{Lampert-MA,Ross-JS}. Charged excitons are a ubiquitous feature of semiconductors, but are difficult to observe in traditional systems -- like GaAs/AlGaAs quantum wells -- due to their small binding energies (1-2 meV) \cite{Kheng-K,Finkelstein-G,Solovyev-VV}. In TMDs, however, both singlet and triplet charged species have been discovered with E$_{\rm{B}}\approx 20-40$ meV \cite{Mak-KF,Ross-JS,Srivastava-A,Lyons-TP,Courtade-E}. In the high carrier density regime, these resonances have been alternatively interpreted as many-body polaron states \cite{Back-P,Sidler-M,Glazov-MM, Efimkin-D1}. 

In analogy to the hydrogen atom, excitons are known to form a Rydberg series of higher energy states \cite{Kazimierczuk-T}. In TMDs, they have been observed through a variety of different optical techniques up to principal quantum number $n$ =11 \cite{Wang-T,Stier-AV,Stier-AV2,Chernikov-A2,Goryca-M,Chen-SY2}. However, even in the presence of excess charge, a corresponding series for the charged exciton has remained elusive. Lack of experimental observations of these states has been thought of analogously to the H$^-$ ion, for which there exists no bound excited state \cite{Hill-RN}. More recent theoretical work \cite{Fey-C,Yan-J,Rau-ARP,Shiau-SY} has detailed scenarios in which these higher $n$ charged excitons could exist, but that work does not match with a series of compelling experimental reports of metastable 2$s$/$3s$ charged excitons in TMDs \cite{Goldstein-T,Liu-E2,Arora-A,Wagner-K}. Significant work remains to reconcile experimental results with theoretical understanding. 

The difficulty in observing these higher $n$ states is two-fold: (I) the weak radiative decay rate of excitons with higher $n$ makes them increasingly dim in typical photoluminescence (PL) measurements \cite{haugkoch} and (II) even once the state is observed optically, further carrier-density and magnetic field dependent measurements are needed to correctly identify the exciton species. In particular, magneto-optical characterization of exciton species has proven to be an important tool for distinguishing between particle configurations \cite{Liu-E4,He-M}. It also gives valuable information about the spin-valley character of each excitation \cite{Li-Z3,Srivastava-A,MacNeill-D,Li-Y,Lyons-TP}, many-body interaction \cite{Back-P,Klein-J,Wang-Z}, and the underlying band structure of the materials themselves \cite{Plechinger-G2,Deilmann-T,Forste-J,Wozniak-T,Robert-C2}.

\begin{figure*}[t]

\includegraphics[width=7in]{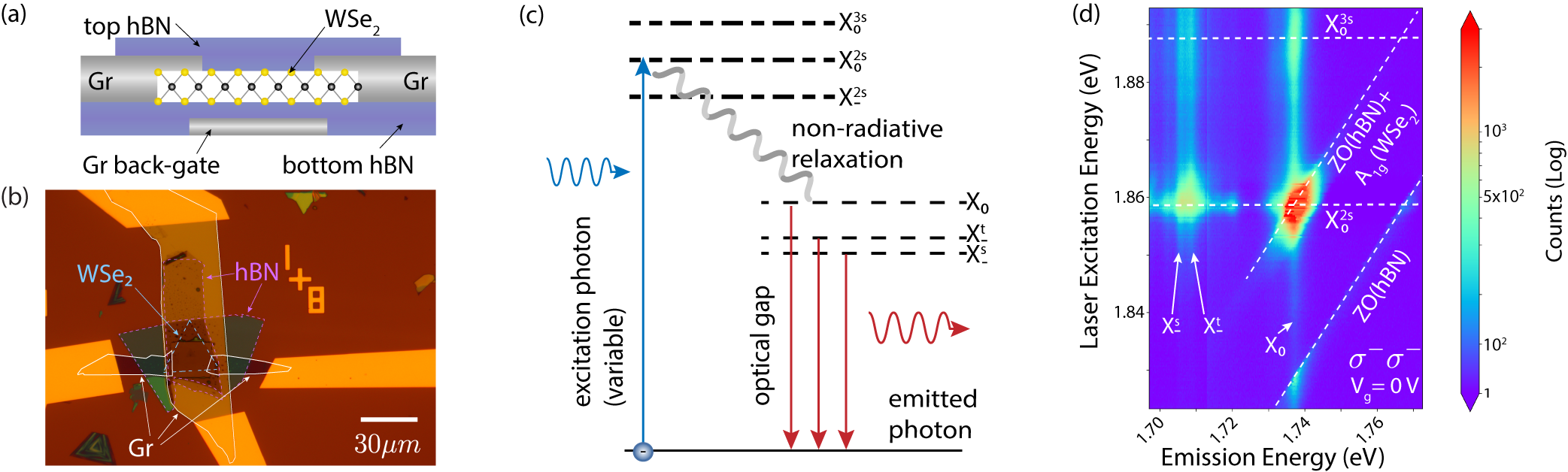}
{\caption{(A) Schematic of the hBN encapsulated WSe$_2$ with graphite (Gr) back gate and electrodes. (B) Optical image of device after full fabrication. The compressed area in the center indicates the region of the sample that underwent nanosqueegeeing. (C) Schematic of the PLE process highlighting the higher $n$ states (e.g. X$_0^{\rm{2s}}$) and emission monitored channels (e.g. X$_0$, X$^{\rm{t}}_-$, or X$^{\rm{s}}_-$). (D) $\sigma^-\sigma^-$ PLE spectra taken at V$_{\rm{g}}$ = 0 V and sample temperature of $<$300 mK. The monitored emission channels are marked with arrows (X$_0$, X$^{\rm{t}}_-$, and X$^{\rm{s}}_-$) \cite{Liu-E,Barbone-M}. Two Raman modes are identified as diagonal dashed lines (ZO(hBN) and ZO(hBN)+A$_{\rm{1g}}$(WSe$_2$)), see Ref. \cite{Suppl,Jin-C,Jadczak-J}. The resonances of the excited states are marked with horizontal dashed lines (X$_0^{\rm{2s}}$, X$_0^{\rm{3s}}$) \cite{Stier-AV2}. 
}\label{fig:fig1}}
\end{figure*}   

In this Letter, we confirm the presence of negatively charged 2$s$-exciton (X$^{\rm{2s}}_-$) in WSe$_2$ via photoluminescence excitation measurements (PLE). In PLE, we monitor the emission from the $1s$ (lowest energy) exciton species while the excitation laser's energy was swept in the energy regime needed to resonantly probe higher $n$ states. This provides a superior signal-to-noise ratio compared to  PL. Additionally, we report on the response of X$^{\rm{2s}}_-$ to an applied magnetic field. We measure the valley dependent Zeeman splitting for both the $2s$ neutral (X$_0^{\rm{2s}}$) and charged (X$^{\rm{2s}}_-$) excitons in the carrier density regime in which they coexist. From this, we extract similar $g$-factors for X$_0^{\rm{2s}}$/X$^{\rm{2s}}_-$, g$_{\rm{X_0^{2s}}}$=-5.20$\pm0.11$ and g$_{\rm{X^{2s}_-}}$=-4.98$\pm0.11$, and discuss the possible physical origins of this result. 



In our experiment, a monolayer of chemical vapor deposition (CVD) grown WSe$_2$ is encapsulated in hBN along with few-layer graphite (Gr) contacts and bottom gate electrode. Encapsulation was performed via the wet capillary action method and interlayer contamination was removed via the nano-squeegee method \cite{Rosenberger-MR} (see Fig.~\ref{fig:fig1}(A) for a schematic of the sample and Fig.~\ref{fig:fig1}(B) for an image of the final device). The full fabrication details are reported in the supplemental material (SM \cite{Suppl}, which includes Refs. \cite{Steinhoff-A,Liu-E3,Molas-MR,Zhang-XX,Kormanyos-A,Manca-M,OHaver-T,Dagupta-PK,Wahab-MF,Ajayi-O,Lin-Y,Zhu-X,LM-fit, Levenberg-K, Chow-C, Serrano-J, Jin-Z,Liu-GB} that are not included in the main text). The joint hBN and Gr encapsulation allows for a high-quality device with electrostatic control over the carriers in the system via the applied gate voltage V$_{\rm{g}}$\cite{Dean-CR}. 

Throughout our work, we utilize PLE to resonantly probe the 2$s$ exciton states. In PLE, the energy of the input photons is varied and when their energy resonantly matches a 2$s$ exciton state, electrons are excited from the valence band to form these excitons (e.g. X$_0^{\rm{2s}}$). There, the $2s$ excitons undergo non-radiative relaxation to a 1$s$ state (e.g. the neutral exciton X$_0$) where they radiatively recombine and emit photons. An illustration of this process is shown in Fig.~\ref{fig:fig1}(C). For these measurements, the excitation beam is generated using a dye laser with a dynamic excitation range of 1.77-1.99 eV. We use a confocal configuration with circular polarization resolution in both excitation and detection. Throughout the text, we denote the excitation/emission polarization in the format $\sigma^{\rm{excitation}}\sigma^{\rm{emission}}$. The sample was placed in a dilution refrigerator equipped with a 12 T superconducting magnet in a Faraday geometry. We estimate that with residual heating from the laser and magnet, the ambient temperature of the sample is $<$300 mK.

Fig.~\ref{fig:fig1}(D) shows a baseline PLE spectrum taken with $\sigma^-\sigma^-$ (-K-valley selective) at V$_{\rm{g}}$ = 0V and B = 0T. We identify the 2$s$ and 3$s$ neutral Rydberg excitons (X$_0^{\rm{2s}}$, X$_0^{\rm{3s}}$) by their binding energies \cite{liu2019magnetophotoluminescence, Chen-SY2} and labeled them with white dashed lines at 1.859 eV and 1.887 eV, respectively. The 1$s$ neutral (X$_0$) exciton's emission channel and the triplet (X$^{\rm{t}}_-$)/singlet (X$^{\rm{s}}_-$) charged excitons' emission channels were identified by their binding energies \cite{Barbone-M, Li-Z4} and PL gate voltage dependence \cite{wang2017probing} (see Ref. \cite{Suppl}).

\begin{figure*}[t]
\includegraphics[width=7in]{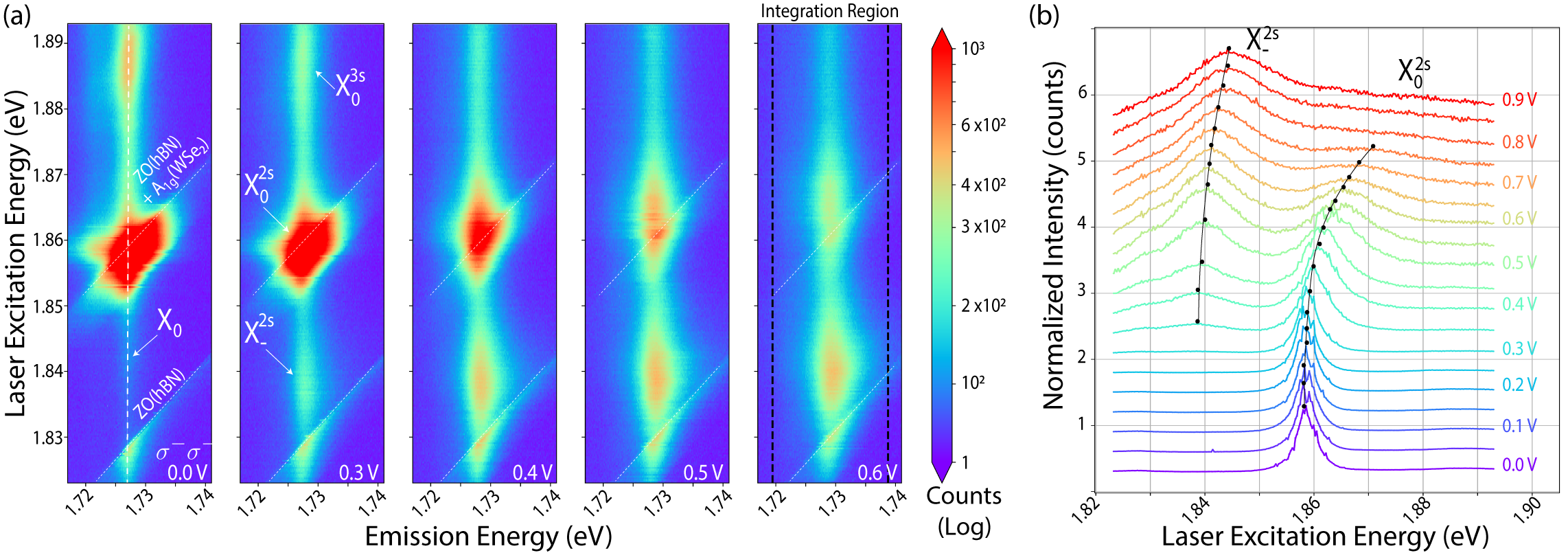}
 
\caption{(A) PLE data with increasing $n_{\rm{e}}$-doping while monitoring the X$_0$ recombination channel in the -K-valley ($\sigma^-\sigma^-$). (B) Waterfall plot of vertical cross-sections from V$_{\rm{g}}$ = 0 - 0.9 V. The integration region is annotated in panel (A). The counts were summed over the emission width for each excitation energy.}
\label{fig:fig2}
\end{figure*}

Next, we tune V$_{\rm{g}}$ to $n_{\rm{e}}$-dope the system and look for signs of an emerging charged 2$s$ exciton in our PLE spectra. Fig.~\ref{fig:fig2} highlights the results of this while monitoring the X$_0$ emission channel;  Fig.~\ref{fig:fig2}(A) shows the full PLE spectra at selected V$_{\rm{g}}$, while Fig.~\ref{fig:fig2}(B) is the integrated vertical cross-section of the emission spectrum around the X$_0$ signal. The integration region used for all gate voltages is denoted in the V$_{\rm{g}}$ = 0.6 V panel of Fig.~\ref{fig:fig2}(A) by the vertical dashed lines. As in Fig.~\ref{fig:fig1}, we identify the resonance at 1.859 eV as X$_0^{\rm{2s}}$. 

\begin{figure*}
\includegraphics[width=7in]{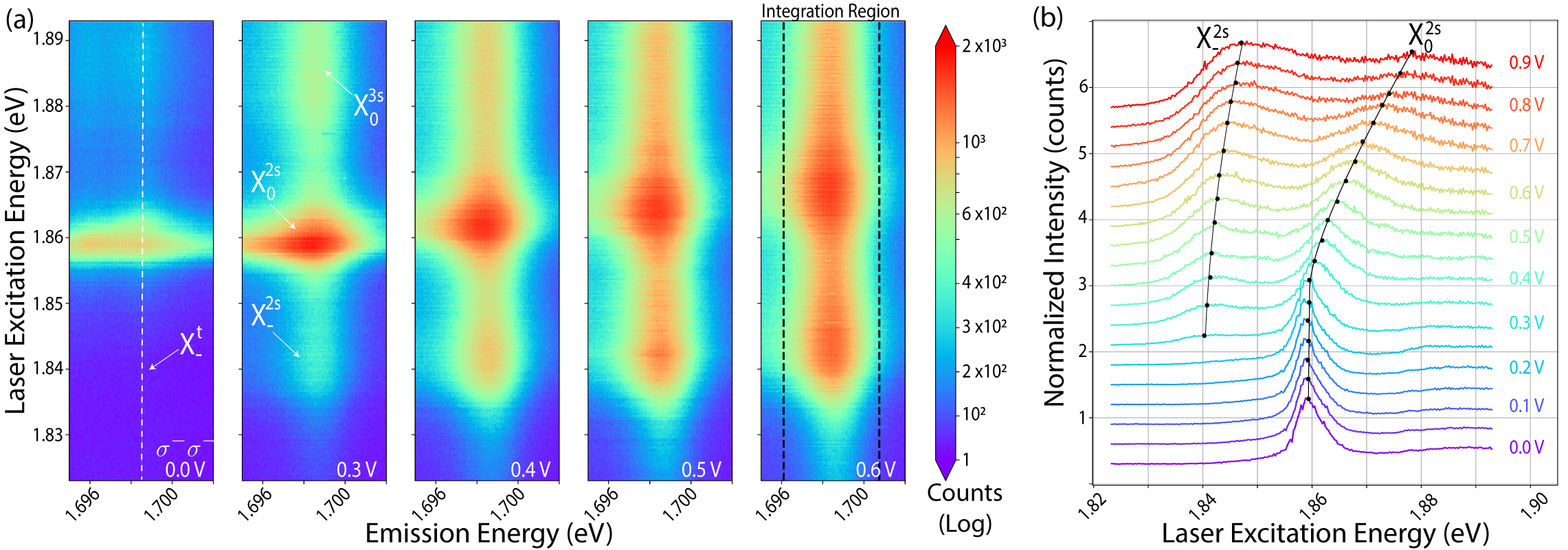}
\caption{(A) PLE data with increasing $n_{\rm{e}}$-doping while monitoring the X$^{\rm{t}}_-$ recombination channel ($\sigma^-\sigma^-$). (B) As in Fig.~\ref{fig:fig2}, the waterfall plot corresponds to vertical cross-sections from V$_{\rm{g}}$ = 0 - 0.9 V.}
\label{fig:fig3}
\end{figure*}

At V$_{\rm{g}}$ = 0.3 V, a lower energy resonance begins to emerge at 1.838 eV. We label this state as the 2$s$ charged exciton X$^{\rm{2s}}_-$ and base this identification on two observations: (I)  V$_{\rm{g}}$ = 0.3 V corresponds to the transition of the sample from charge neutrality to $n_e$-doped and the emergence of the negatively charged 1$s$ excitons X$^{\rm{t}}_-$/X$^{\rm{s}}_-$ (see Ref. \cite{Suppl} for $1s$ PL data). The X$^{\rm{2s}}_-$ resonance displays a similar onset at V$_{\rm{g}}$ = 0.3 V indicating a similar negative charge character. (II) When the X$^{\rm{2s}}_-$ resonance first appears at $V_g$=0.3 V,  we find that $\Delta E_{\rm{(X_0^{\rm{2s}}-X^{\rm{2s}}_-)}}$ = 21 meV while $\Delta E_{\rm{(X_0-X_-^{\rm{t}})}}$ = 29 meV and $\Delta E_{\rm{(X_0-X_-^{\rm{s}})}}$ = 35 meV. This reduction indicates that the $2s$ charged exciton is less tightly bound than its $1s$ state counterpart. This is in accordance with other observations in the literature \cite{Wagner-K,Arora-A,Goldstein-T,Liu-E2} and consistent with the fact that Rydberg states display a reduction in relative binding energy with each increasing $n$. 

Since the $2s$ charged exciton is expected to be a doublet, as observed for the $1s$ charged excitons, the extracted position of X$^{\rm{2s}}_-$ is an average. X$_-^{\rm{t}}$ and X$_-^{\rm{s}}$ have a narrow linewidth and a strong intervalley exchange interaction that splits them ($\approx$ 6 meV \cite{Yu-H, Courtade-E}) which allows us to spectrally resolve them. However, the broadness of the 2$s$ states combined with a reduced intervalley exchange energy (theoretically predicted to be $\approx$ 1 meV \cite{Arora-A,Yu-H}), prevents us from resolving the doublet of the $2s$ charged exciton. There is, however, indication of the two states in the asymmetric lineshape of the X$^{\rm{2s}}_-$ peak (see Ref. \cite{Suppl}). 

\begin{figure*} 
\includegraphics[width=7in]{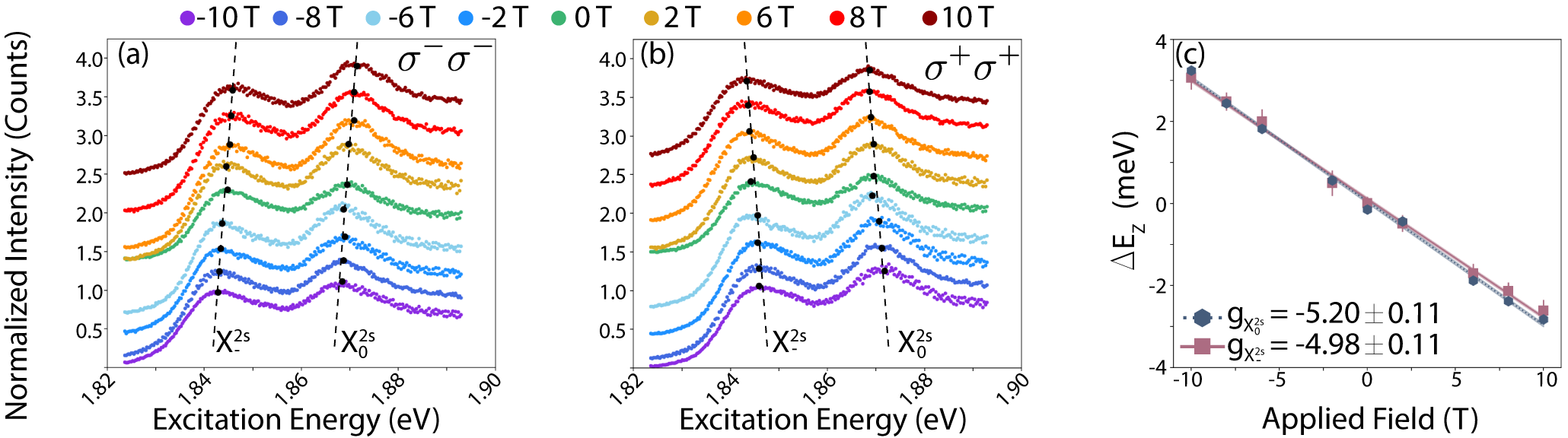}
\caption{Vertical cross-sections from the X$^{\rm{t}}_-$ emission channel as a function of field for (A) ($\sigma^-\sigma^-$) and (B) ($\sigma^+\sigma^+$) marked with the corresponding peak positions (black dots) for the X$_0^{\rm{2s}}$ and X$^{\rm{2s}}_-$ states from fitting with the dashed line serving as a guide to the eye. (C) Extracted $g$-factors for X$_0^{\rm{2s}}$ and X$^{\rm{2s}}_-$ states. The thickness of the fit line in panel (C) corresponds to the error in the fit.}
\label{fig:fig4}
\end{figure*} 

In Fig.~\ref{fig:fig2}(B), we see the spectral dependence of X$^{\rm{2s}}_0$ and X$^{\rm{2s}}_-$ with carrier density. As the $n_{\rm{e}}$-doping increases with increasing gate voltage, the X$_0^{\rm{2s}}$ resonance broadens, decreases in intensity, and spectrally blueshifts. The broadening and loss of spectral intensity are consistent with more rapid decoherence from interaction with the Fermi sea. The blueshift results from the competing effects of band gap and binding energy renormalization due to decreased $\rm{e}^-\rm{-}\rm{e}^-$ and $\rm{e}^-\rm{-}\rm{h}^+$ interaction from screening by the Fermi sea \cite{Chernikov-A,Roch-JG,Wagner-K}.

In contrast, X$^{\rm{2s}}_-$ peak grows in intensity and experiences minimal spectral drift with increased carrier density. In the case of a three-body quasiparticle, one expects a redshift that is linearly dependent on the charge concentration in the system resulting from momentum conservation \cite{Chernikov-A,Mak-KF,Efimkin-D1}. This competes with the effects of band gap and binding energy renormalization previously discussed for the neutral excitons that favor a blueshift \cite{Chernikov-A}, and leads to the minimal spectral drift observed. Both the increase in intensity and small spectral shift are consistent with the behavior of $1s$ and $2s$ charged excitons previously observed \cite{Wagner-K, Chernikov-A,Roch-JG}.

Since X$^{\rm{2s}}_-$ emerges in the $n_{\rm{e}}$-doped regime, we expect X$^{\rm{t}}_-$ and X$^{\rm{s}}_-$ to be the most prominent emission channels for $2s$ exciton species (see Ref. \cite{Suppl}). To verify this, we monitor the X$^{\rm{t}}_-$ emission channel in a similar manner to X$_0$ and show the results as a function of V$_{\rm{g}}$ in Fig.~\ref{fig:fig3} (the results for X$^{\rm{s}}_-$ can be found in Ref. \cite{Suppl}). We confirm that the behavior (spectral position, shift with gate, etc.) of X$_0^{\rm{2s}}$ and X$^{\rm{2s}}_-$ is independent of the monitored decay channel. 


We turn our attention to extracting the behavior of the X$_0^{\rm{2s}}$ and X$^{\rm{2s}}_-$ with applied magnetic field. We chose to take the data at V$_{\rm{g}}=0.6$ V because both the neutral and charged exciton have similar intensity. Integrated vertical cross-sections of the X$^{\rm{t}}_-$ emission channel presented in Fig.~\ref{fig:fig4} (A)/(B) show the response of the -K($\sigma^-\sigma^-$)/+K($\sigma^+\sigma^+$) valleys, respectively, with magnetic field. The extracted peak centers from fitting are marked with black dots. Applying a magnetic field breaks the time-reversal symmetry in the system, and results in a red(blue) shift with positive field for the +K(-K) valley and vice versa with applied negative field \cite{Srivastava-A,Back-P}. 

Using the definition for the Zeeman splitting in terms of polarization components, $\Delta E_Z = E^{\sigma^+\sigma^+} - E^{\sigma^-\sigma^-}= g\mu_{\rm{B}}B$, we fit a linear model to our data and extract a $g$-factor of -5.20$\pm0.11$ and -4.98$\pm0.11$ for X$_0^{\rm{2s}}$ and X$^{\rm{2s}}_-$, respectively. This fit and extracted difference is shown in Fig.~\ref{fig:fig4}(C). Results that agreed within experimental error were found for both  X$^{\rm{2s}}_0$ and X$^{\rm{2s}}_-$ for a similar analysis of the X$^{\rm{s}}_-$ emission channel (see Ref. \cite{Suppl}).

Frequently, a single-particle model is used to interpret the $g$-factor for $1s$ excitons. In this model, the contributions to the Zeeman splitting are defined as $\Delta E_{\rm{Z}} = -\vec{\mu}\cdot\vec{B}$. The magnetic moment $\vec{\mu}$ is composed of additive terms for the orbital and spin contributions (intracellular components $\mu_{\rm{O}},\mu_{\rm{S}}$) along with a correction for the effects of the finite Berry curvature in the system (intercellular component $\mu_{\rm{V}}$) \cite{MacNeill-D,Li-Y,Forste-J,Koperski-M,Yu-H} in each relevant band. Within this interpretation, we expect $g_{\rm{X_0}} \approx -4.4$ and $-11$ $\lessapprox g_{\rm{X_-^{t/s}}}\lessapprox -4 $  (depending on the method used to calculate $\mu_{\rm{V}}$, and whether the doublet is resolved \cite{Srivastava-A, Lyons-TP,Liu-E}). 

To serve as a reference point between the literature and our 2$s$ results, we also extracted the $g$-factors for $\rm{X_0}$ and $\rm{X^{t/s}_-}$. These values are g$_{\rm{X_0}}$ = -4.22$\pm$0.04, g$_{\rm{X_-^{\rm{t}}}}$ = -4.12$\pm$0.04, and g$_{\rm{X_-^{\rm{s}}}}$ = -3.86$\pm$0.05 in our system at V$_{\rm{g}}$=0.6 V. They are consistent with the results from the single particle interpretation, but highlight a distinct increase in our 2$s$ $g$-factors with respect to the corresponding $1s$ states. We discuss two possible contributions to this enhancement. 

(I) Enhancement of the $g$-factor for the 2$s$ neutral exciton has been observed in magnetic Rydberg measurements in both intrinsic and electrostatically neutral samples \cite{Chen-SY2,Wang-T,Goryca-M}. Since the observation in neutral samples rules out doping effects, the divergence from $g_{\rm{X_0}} \approx -4.4$ has been attributed to enhanced intercellular contributions arising from the increased $k$-space localization of the wavefunctions with each subsequent $n$ \cite{Chen-SY2}. Extending this technique to charged excitons gives an intercellular component that \emph{decreases} as the Bohr radius \emph{increases}. This is compounded by an increased $k$-space localization of the charged exciton (see Ref. \cite{Suppl}). While this model could explain the results for X$_0^{\rm{2s}}$ it would underestimate the $g$-factor for X$^{\rm{2s}}_-$.

(II) A second possibility is the onset of many-body interaction (polaron picture) between the excitons and the emerging Fermi sea from electrostatic gating. Many-body interactions are expected to be very favorable in WSe$_2$ which has a Wigner-Seitz radius greater than 1 even at extremely high densities \cite{Wang-Z, DasSarma-S}. The interaction strength will vary with the Fermi sea's population and the Bohr radius, and induce Fermi sea polarization. Carrier dependent enhancement of the $g$-factor in TMDs has been documented for many materials/quasiparticles, with the strength of enhancement dictated by the degree of the induced Fermi sea polarization \cite{Wang-Z,Klein-J,Back-P,Li-Z3}. 

In the many-body picture, it has been observed that as doping levels are varied there is a convergence of the $g$-factor between competing quasiparticles (e.g. X$_0$ and X$^-$) in regions in which they coexist. In analogy to the Kondo effect, the impurity (exciton) is dressed with either an attractive or repulsive interaction with the Fermi sea. As carrier density increases, the state dressing will become more similar for all exciton species -- regardless of the type of interaction -- resulting in a convergence of the $g$-factors \cite{Klein-J} for X$_0$-like and X$_-$-like excitons. Such behavior is not expected to be limited to the $1s$ state excitons and can explain the convergence of our extracted values of $g$ for the X$_0^{\rm{2s}}$ and X$^{\rm{2s}}_-$ within experimental error.

Our results serve as the first marker in mapping the behavior of the $2s$ charged state, X$^{\rm{2s}}_-$, with magnetic field in TMDs. Experimental quantification of the $g$-factor serves as another physical benchmark for future theoretical models of stable 2$s$ charged states. Additionally, the stability of the X$^{\rm{2s}}_-$ state offers a possible medium for studying the cross-over from exciton Rydberg physics to the quantum Hall regime for charged species at high magnetic fields. Recent work by Klein \emph{et al.} used carrier density dependent $g$-factor measurements to demonstrate tunable many-body physics through all 1$s$ exciton species in MoS$_2$ \cite{Klein-J}. Our initial results indicate that it would be possible to produce this type of map for 2$s$ species with access to higher magnetic fields and devices with larger dynamic carrier density range. This opens up a unique opportunity to study many-body interactions in higher energy exciton species that is generally limited in traditional semiconductors systems with smaller exciton binding -- like GaAs quantum wells.

\begin{acknowledgments}
The authors acknowledge fruitful discussions with A. Friedman,  T. O'Haver, G. Solomon, A. Srivastava, and E. Waks. The work at Maryland was supported by AFOSR FA95502010223, NSF PHY1820938, and NSF DMR-2019444, ARL W911NF1920181, ARO W911NF2010232, Simons and Minta Martin Foundations. J.C.S. acknowledges additional support from the ARCS Scholar program. The work done at NRL was supported by core programs funding.
\end{acknowledgments}

\end{document}


\title{Magneto-Optical Measurements of the Negatively Charged 2$s$ Exciton in WSe$_2$: Supplementary Material}
\author{{{J.C.Sell$^*$, J.R.Vannucci\thanks{These authors contributed equally}}, D.G. Su\'{a}rez-Forero, B. Cao, D.W. Session, H.-J. Chuang, K.M. McCreary, M.R. Rosenberger, B.T. Jonker, S. Mittal, M. Hafezi}}
\maketitle
\begin{figure*}[h]
\includegraphics[width=\linewidth]{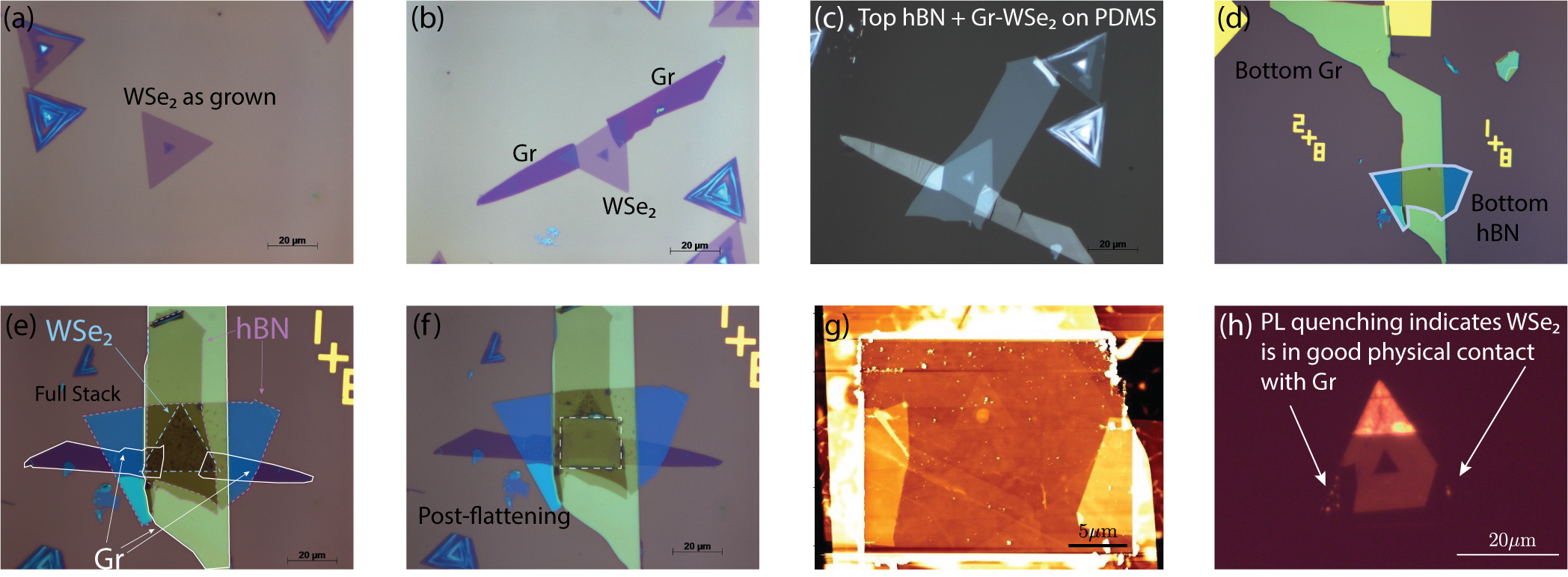}
{\caption{More detailed images from the fabrication process showing (A) the as-grown WSe$_2$, (B)/(C)/(D) graphite leads/gate, (E) full hBN encapsulation, (F) fully encapsulated stack after nano-squeegee process (highlighted with dashed box), (G) an AFM image of the nano-squeegee region highlighting the removal of interlayer impurities from the stacking process, and (H) a PL image of the WSe$_2$ monolayer and graphite leads, with the quenching of emission at the contact points indicating good physical contact. 
}\label{fig:supp:encap}}
\end{figure*} 
\section{Fabrication Details}
\emph{CVD growth of WSe$_2$ --} Chemical vapor deposition (CVD) synthesis of WSe$_2$ was performed in a two inch quartz tube furnace on SiO$_2$/Si (275 nm oxide) substrates. Prior to use, all SiO$_2$/Si substrates were cleaned in acetone, isopropanol (IPA), and Piranha etch (H$_2$SO$_4$ + H$_2$O$_2$) then thoroughly rinsed in deionized water. At the center of the furnace was positioned a quartz boat containing approximately 1g of WO$_3$ powder (Alfa Aesar 99.999$\%$). Two SiO$_2$/Si wafers were positioned face-down, directly above the oxide precursor. A separate quartz boat containing approximately 500  mg selenium powder (Alfa Aesar 99.999$\%$) was placed upstream, outside the furnace-heating zone. The upstream SiO$_2$/Si wafer contained perylene- 3,4,9,10-tetracarboxylic acid tetrapotassium salt (PTAS) seeding molecules to promote lateral growth, while the downstream substrate was untreated. Pure argon (65 sccm) was used while the furnace ramped to the target temperature. Upon reaching the target temperature of 825 \textdegree C, 10 sccm H$_2$ was added to the Ar flow and maintained throughout the 10 minutes soak and subsequent cooling to room temperature. 

\emph{Heterostructure construction --} Mechanical transfer from the deposition substrate and van der Waals heterostructure construction was performed using a wet capillary technique outlined in the experimental methods section of Ref. \cite{Rosenberger-MR}. The polydimethylsiloxane (PDMS) used in this process was made from a commercially available SYLGARD 184 silicone elastomer kit. To prepare the PDMS mixture, we thoroughly mix Silicone Elastomer and curing agent with a weight ratio of 10:1 followed by a debubbling process under rough vacuum for 30 minutes. This mixture was spin coated on a silicon wafer with a spin rate of 350 rpm for 30 s, then cured at 80 °C for 20 minutes on a hot plate. The resultant PDMS is easily peeled off the silicon wafer for use. 

The top and bottom hBN in the heterostructure are 12 nm and 15 nm thick, respectively, as measured by AFM. After the heterostructure was complete, interlayer interfacial contamination was removed via the nano-squeegee method \cite{Rosenberger-MR}. Fig.~\ref{fig:supp:encap} shows more detailed images from the encapsulation process. 

\emph{Electrical contact fabrication --} After the full heterostructure was placed onto the SiO$_2$/Si substrate, electron beam lithography (EBL) was used to define leads to the graphite contacts and back gate. For this, a bilayer of poly (methyl methacrylate) (PMMA) 950 A4 was spun onto the sample and baked at 185 \textdegree C for 5 and 10 minutes, respectively. The patterning was performed on an 100kV system. Post patterning and development, 3 nm Cr/70 nm Au contacts were deposited using thermal evaporation. Excess metal was removed using a standard solvent liftoff procedure in a bath of room temperature acetone for 1 hour. 

\section{Optical Configuration}
\begin{figure*}[tb]
\includegraphics[width=\linewidth]{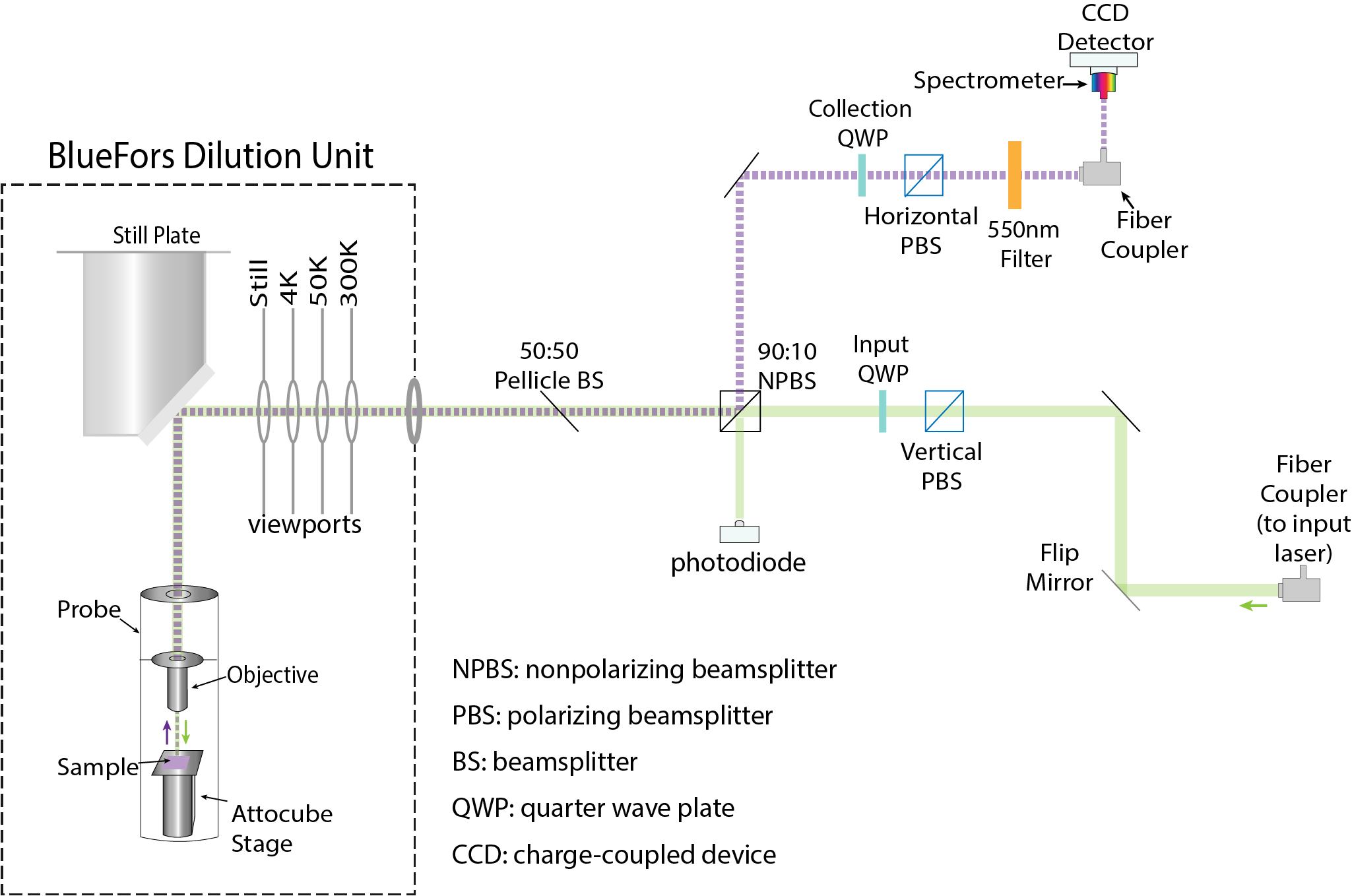}
{\caption{Optical diagram depicting the excitation and emission collection scheme used for photoluminescence and photoluminescence excitation measurements. 
}\label{fig:supp:optical}}
\end{figure*} 

There were two different excitation sources used throughout the main manuscript and the supplemental information. For the photoluminescence excitation (PLE) measurements in the main text, a dye laser with  4-(Dicyanomethylene)-2-methyl-6-(4-dimethylaminostyryl)-4H-pyran (DCM) dye and pumped with a 2.33 eV (532 nm) green laser was used to access a dynamic range from 1.92-1.75 eV. The dye laser was also utilized for PL measurements of the 1$s$ state, while for PL measurements accessing both the $n$=1,2 states a 2.33 eV green diode laser was used.  The laser's power was monitored and stabilized via a PID loop at a constant power density of approximately 6.8 kW/cm$^2$ throughout all our measurements. The illumination spot size is estimated to be approximately 1 $\mu$m. As with the PLE measurements in the main text, all supplemental measurements were also performed in the confocal configuration and Faraday geometry. An illustration of the optical scheme is shown in Fig.~\ref{fig:supp:optical}. 

\section{Electrostatic Gate Mapping of the 1$s$ Exciton Species}
\begin{figure*}[t]
\includegraphics[width=6in]{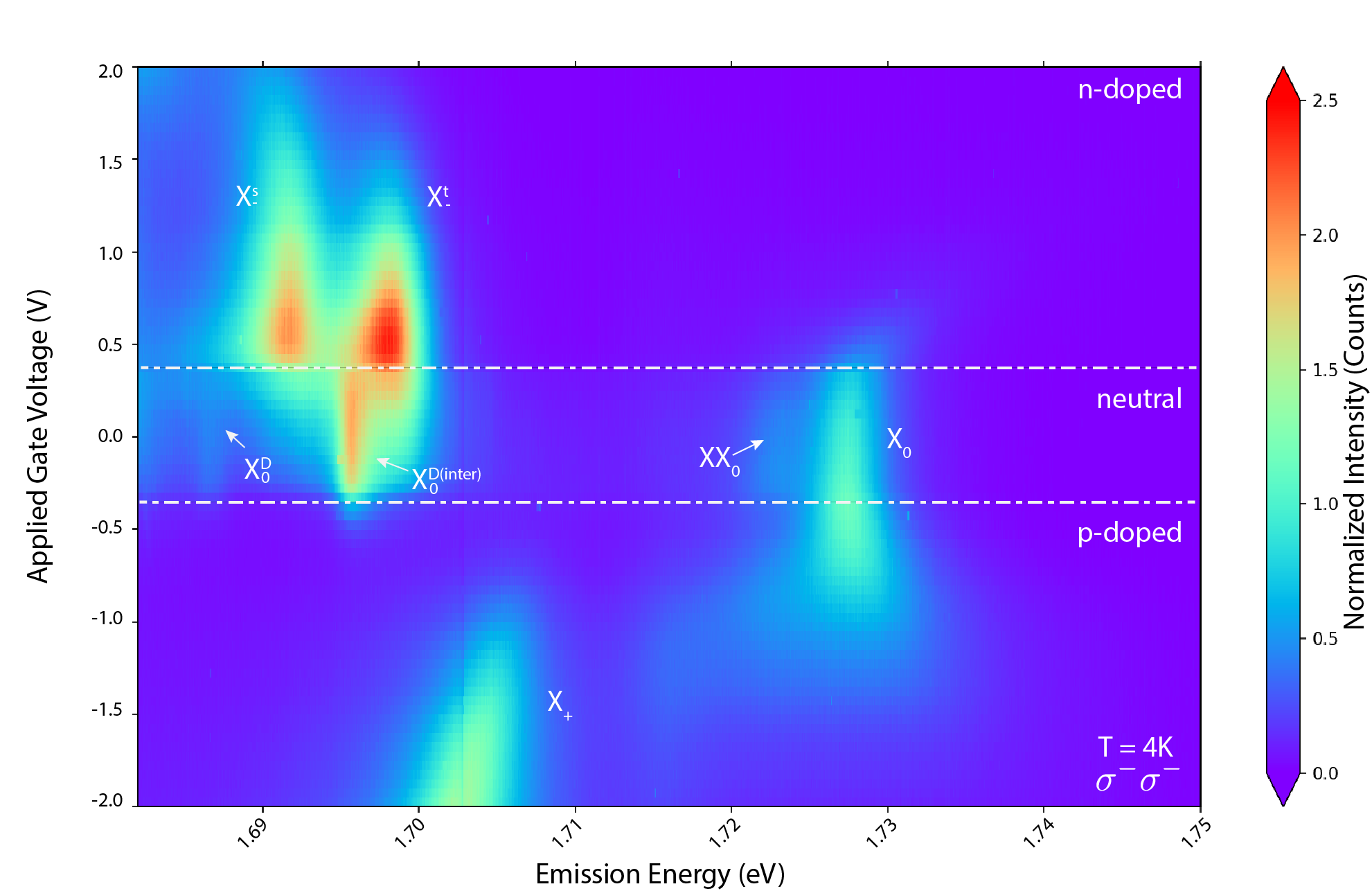}
{\caption{PL of the 1$s$ exciton species as a function of gate voltage. Here, there are three regions marked based on their charge character: $n$-doped, neutral, and $p$-doped. Exciton species are noted. Data collected while the sample was illuminated with an excitation energy of 1.95 eV.
}\label{fig:supp:gate}}
\end{figure*}

An initial gate map was produced to identify different doping regions. For this measurement, polarization-resolved PL was collected as a function of gate voltage from -2V to 2V.  The results presented in Fig.~\ref{fig:supp:gate} are from a mapping using $\sigma^-\sigma^-$, but $\sigma^+\sigma^+$ was also collected and found to produce identical results. The spectra is normalized to the 1$s$ exciton intensity at V$_{\rm{g}}$ = 0V. 

In this measurement, we identify three charge regimes: neutral, $n$-doped, and $p$-doped. In the neutral regime we identify the 1$s$ exciton X$_0$\cite{Liu-E,Srivastava-A}, the neutral biexciton cluster XX$_0$\cite{Steinhoff-A}, the intervalley momentum-dark exciton X$^{\rm{D(inter)}}_{\rm{0}}$\cite{Liu-E3}, and the intravalley spin-dark exciton X$^{\rm{D}}_{\rm{0}}$ \cite{Liu-E,Molas-MR,Zhang-XX}.

In the charge neutral regime, the Fermi level remains in the band gap and only localized disorder states increase in occupation. These band gap states have little influence on the overall electronic properties of the sample outside of allowing for a small probability that charged exciton states will form through coupling to these localized electrons. 

The sample enters the $n$-doped regime when carriers start to populate the lower conduction bands at V$_{\rm{g}} \approx 0.3$V. This region is host to the triplet X$_-^{\rm{t}}$ and singlet X$_-^{\rm{s}}$ charged excitons \cite{Liu-E,Barbone-M,Lyons-TP}. Their appearance has a significant impact on the exciton emission spectrum from the sample. As the Fermi level crosses the lowest conduction band, both the $\mathrm X_-^{s}$ and $\mathrm X_-^{t}$ quickly increase in brightness while the neutral exciton starts to rapidly blue-shift and dim.

On the other side of charge neutrality, the sample enters the $p$-doped regime at V$_{\rm{g}} \approx -0.3$V. Here, the dominant excitation is the positively charged exciton X$_+$\cite{Liu-E}. 

Exciton species labeled in Fig.~\ref{fig:supp:gate} were identified based on the doping regime and their spacing from the fundamental X$_0$ excitation at onset gate voltage. As was briefly mentioned in the main text, the charged exciton peaks are sometimes interpreted as polarons \cite{Back-P,Sidler-M}. This remains an active area of research at the time of this manuscript with recent work showing that polaron and charged exciton interpretations converge in their predictions at low-to-intermediate carrier density \cite{Glazov-MM}. These results highlight the difficulty in drawing a dividing line between the two interpretations. For simplicity we refer to them as charged excitons throughout the text. 

\section{Extraction of Carrier Concentration}
\begin{figure*}
\includegraphics[width=5.8in]{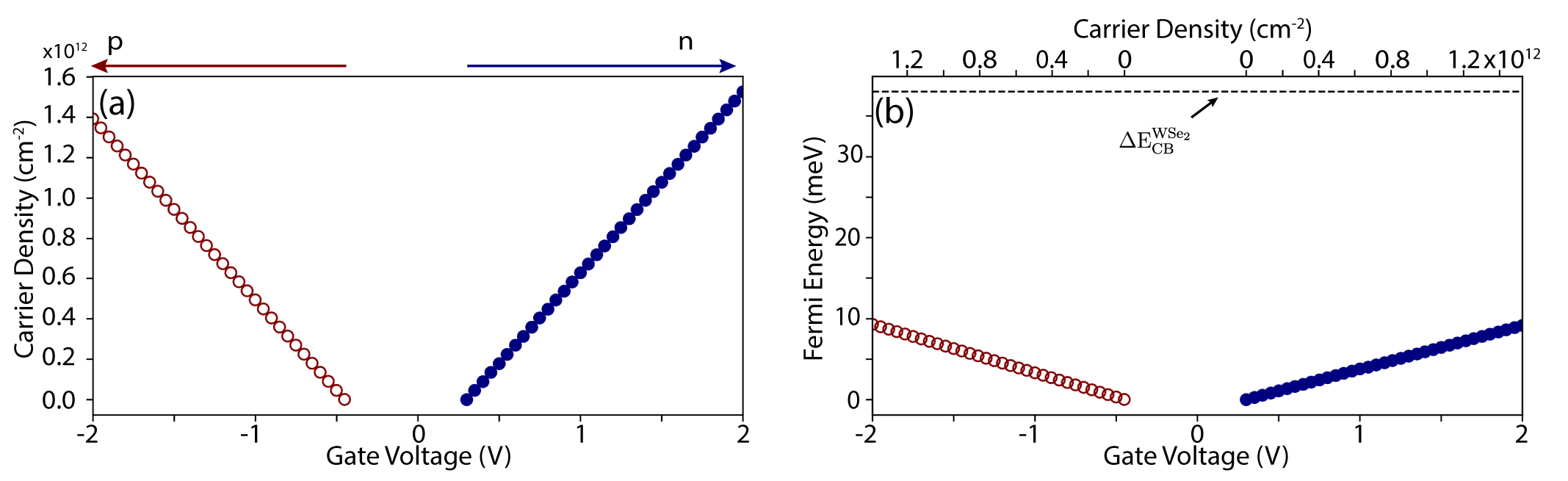}
{\caption{(A) Extracted carrier density as a function of applied gate voltage in a capacitive model. (B) Calculated Fermi energy from carrier dependence with gate voltage. The energy corresponding to the conduction band spin-splitting is noted.
}\label{fig:supp:carrier}}
\end{figure*} 
Using a standard capacitive model, we estimate the carrier density as a function of gate voltage via the following relation \cite{Back-P}, 
\begin{equation}
    \Delta n_e = \frac{\Delta V\epsilon_0\epsilon}{eL}
\label{eqn:carriers}
\end{equation}
Here, $\epsilon_0$ is the permittivity of free space, $e$ is the elementary charge, $L$ is the thickness of the dieletric spacer, and $\epsilon$ is the relative permittivity of the dieletric spacer. In our case, the dieletric spacer is the hBN between the graphite back gate and our sample, which is estimated to be 15 nm thick. The thickness was measured using atomic force microscopy (AFM) and is expected to have a margin of error of approximately 5\%. The average dielectric background for the hBN is given as $\epsilon_\perp\approx2.5$ \cite{Wang-Z2}. We estimate the charge neutrality region based on the extension of neutral states in Fig.~\ref{fig:supp:gate}, which extends from V$_{\rm{g}}$=-0.3 to 0.3 V. Fig.~\ref{fig:supp:carrier}(A) shows the results of applying Eqn. \ref{eqn:carriers} with our system parameters.

We can use the carrier density to extract an approximate associated Fermi level in the sample. Here, we estimate the Fermi energy based on the carrier density as:
\begin{equation}
    E_{\rm{F}} = \frac{n}{\rho(E)} = \frac{n}{m^*/(\pi\hbar^2)},
\end{equation}
with $\rho(E) = m^*/(\pi\hbar^2)$ as the density of available states when the the Fermi level is below the upper conduction band. We take the effective mass of the electron m$^*_{\rm{e}}$ = 0.4$m_{\rm{e}}$ \cite{Liu-GB} and the effective mass of the hole to be m$^*_{\rm{h}}$=0.36m$_{\rm{e}}$ \cite{Kormanyos-A}. WS$\rm{e_2}$ transitions from a degenerate to a non-degenerate state as the Fermi level enters the upper conduction band. This results in the density of states becoming $\rho(E) = m^*/(2\pi\hbar^2)$. The Fermi energy is measured from the bottom of the lower conduction band at B=0T. Using these physical parameters, we generate \ref{fig:supp:carrier}(B) and indicate that the splitting of the upper and lower conduction bands for our system is $\Delta E_{\rm{CB}}^{\rm{WSe_2}}$ = 38 meV\cite{Liu-GB}.

We estimate that the carrier concentration in the system is $n_{\rm{e}}$ $\approx$ 2.7$\times 10^{11}$cm$^{-2}$  at V$_{\rm{g}}$=0.6 V based on the above capacitive model of our device. Based on extracted $g$-factors, magnetic field range, and carrier concentration we expect the carriers in the system to only be partially valley polarized.
%
\section{2$s$ Photoluminescence}
\begin{figure*}[tb]
\includegraphics[width=\linewidth]{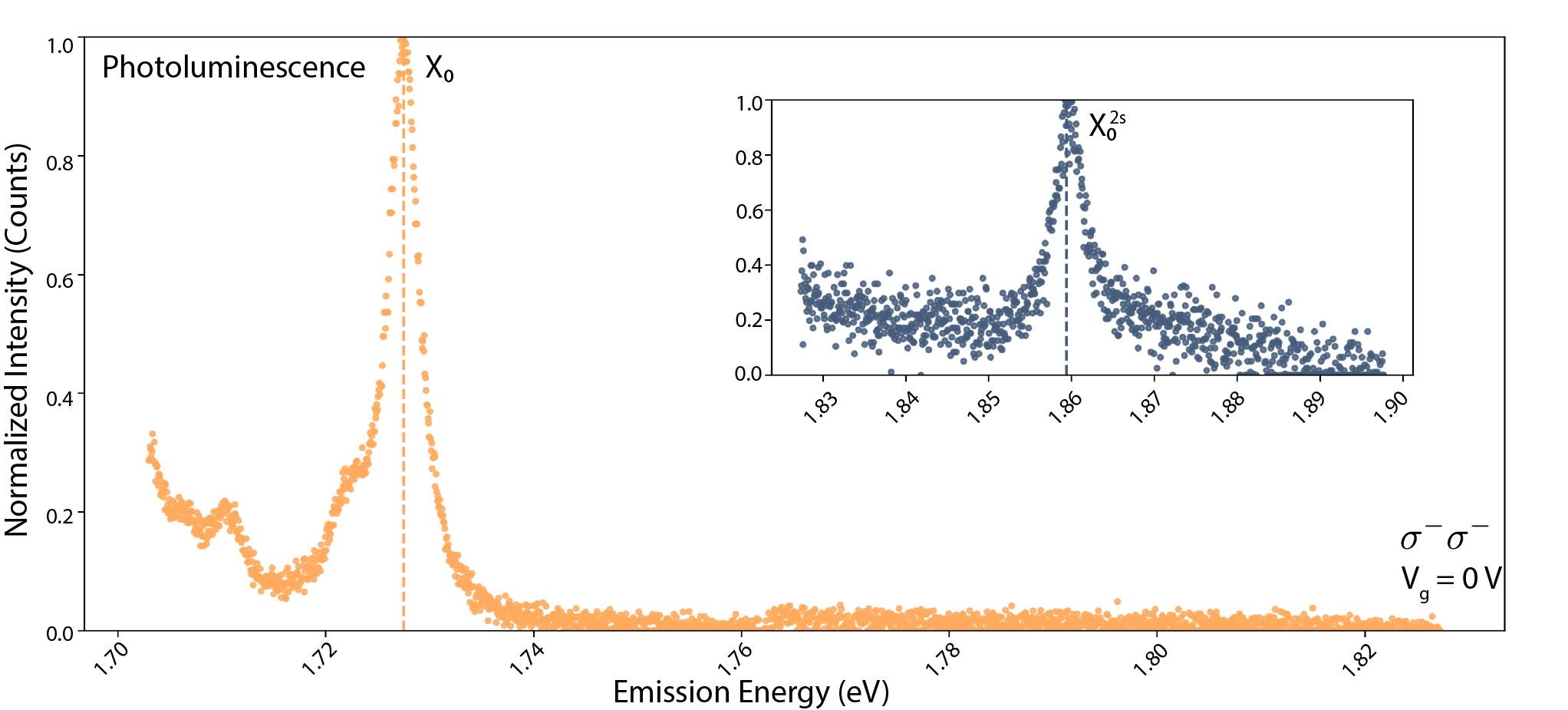}
\caption{Comparison of the PL signal for X$_0$ and X$_0^{\rm{2s}}$ for PL measurements with $\rm{E_{\rm{ex}}}$ = 2.33 eV. Both spectra are normalized to the emission maximum of their respective exciton lines and the spectral position of each is marked with a dashed line.}
\label{fig:supp-2s}
\end{figure*} 
To collect PL from both the 1$s$ and 2$s$ excitons, we used a green laser ($\rm{E_{\rm{ex}}}$ = 2.33 eV) and hold the sample at V$_{\rm{g}}$ = 0V. The results of this measurement are shown in Fig.~\ref{fig:supp-2s}. We find that E$_{\rm{X_0}}^{\rm{PL}}$ = 1.728 eV and E$_{\rm{{X_0^{\rm{2s}}}}}^{\rm{PL}}$ = 1.859 eV. This is comparable to the values extracted from PLE, E$_{\rm{X_0}}^{\rm{PLE}}$ = 1.727 eV and E$_{\rm{{X_0^{\rm{2s}}}}}^{\rm{PLE}}$ = 1.859 eV. In either case, we find the $\Delta E_{\rm{1s-2s}} \approx$ 132 meV. This is comparable to a spacing of $\Delta E_{\rm{1s-2s}} =$ 130 meV found for WSe$_2$ in the literature \cite{Stier-AV2,Manca-M}.

We note that while the excitation energy is high enough to probe the 3$s$ exciton as well, we do not observe this state in our PL measurement. We attribute this to increased noise (evident in Fig.~\ref{fig:supp-2s} around the X$_0^{\rm{2s}}$ region) and increasingly low oscillator strength for the higher $n$ states \cite{haugkoch}. 

In Fig.~1(D) of the main text, we are able to measure a resonance attributed to X$_0^{\rm{3s}}$ at E$_{\rm{{X_0^{\rm{3s}}}}}^{\rm{PLE}}$ = 1.884 eV. Its offset of $\Delta E_{\rm{2s-3s}} \approx$ 25 meV is comparable to the $\Delta E_{\rm{2s-3s}} =$ 22 meV measured in previous reports \cite{Stier-AV2}.

\section{$\rm{X_-^{2s}}$ Photoluminescence}
\begin{figure*}[tb]
\includegraphics[width=\linewidth]{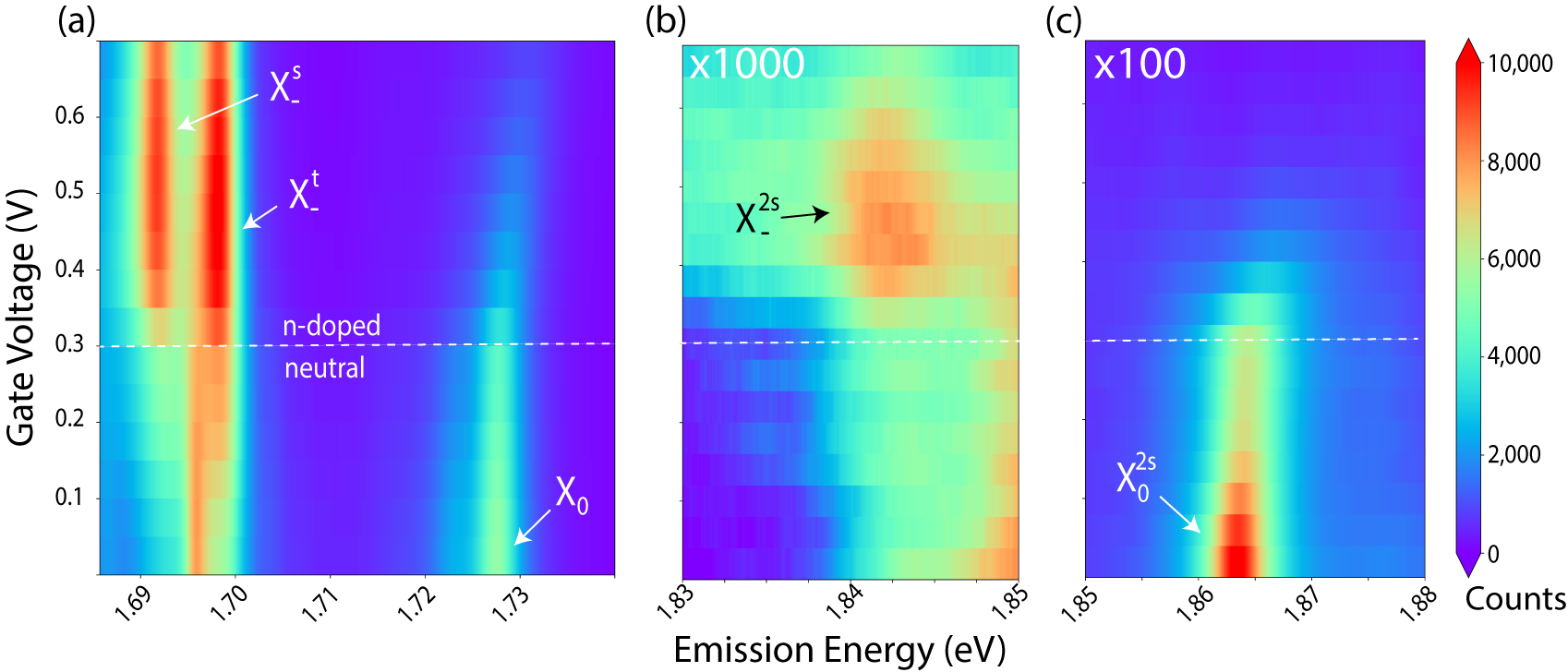}
\caption{Comparison of the PL signal for X$_0$, $\rm{X_-^{2s}}$, and X$_0^{\rm{2s}}$ for PL measurements with gate dependence. All spectra have been normalized to the same color bar with the scale factor highlighted in the top left of each plot. (A) The 1s exciton species were collected while the sample was illuminated with an excitation energy of 1.95 eV (633 nm). (B/C) The 2$s$ exciton species were collected while the sample was illuminated with an excitation energy of 2.33 eV. The sample was illuminated with linearly polarized light and the collection was polarization resolved for $\sigma^-$.}
\label{fig:supp-2s-}
\end{figure*} 

A similar measurement technique to what was outlined in the previous section was used to collect the PL response of the sample as the gate voltage was swept into the $n$-doped regime. In addition to collecting the PL recombination energies from the 1$s$ and 2$s$ neutral excitons, we can also see the emergence of the 2$s$ charged exciton at V$_{\rm{g}}$ $\approx +0.3$V.  The data presented in Fig.~\ref{fig:supp-2s-} shows the emergence of the X$_-^{\rm{2s}}$ resonance at E$_{\rm{X_-^{2s}}}^{\rm{PL}}$ = 1.841 eV. This is in close agreement with our reported value of E$_{\rm{X_-^{2s}}}^{\rm{PLE}}$ = 1.838 eV from the PLE.

Although we are able to spectrally resolve the X$_0^{\rm{2s}}$ and X$_-^{\rm{2s}}$ through PL in Fig.~\ref{fig:supp-2s-}(B/C), the number of counts from the radiative recombination of X$_0^{\rm{2s}}$ is over two-orders of magnitude smaller than for X$_0$. The PL contrast is even higher between the X$_-^{\rm{2s}}$ and X$_-^{\rm{t/s}}$. Hence, in order to reliably measure the spectral profile of the higher energy excitons, we moved to PLE. 

%
\section{X$_-^{\rm{2s}}$ Emission Channel Evolution with Gate}
\begin{figure*}[tb]
\includegraphics[width=\linewidth]{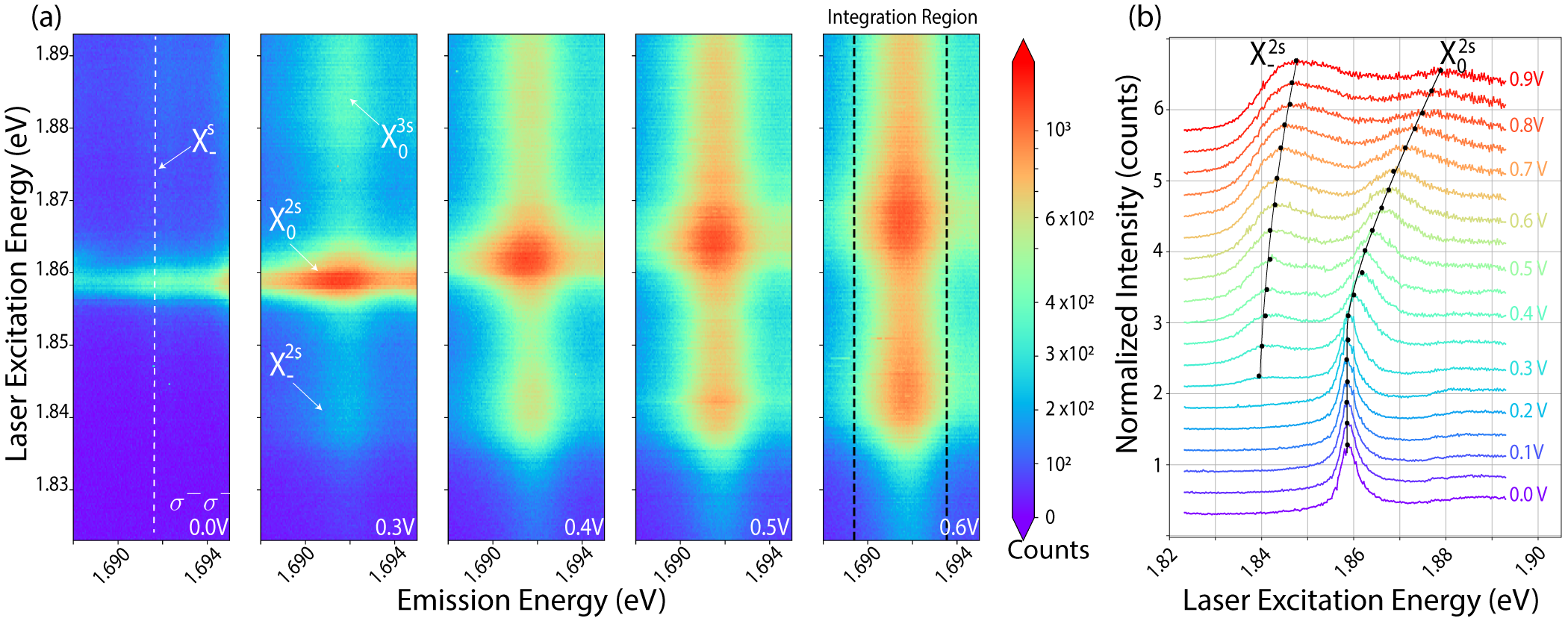}
\caption{(A) PLE data with increasing $n_{\rm{e}}$-doping while monitoring the X$^{\rm{s}}_-$ recombination channel in the -K-valley ($\sigma^-\sigma^-$). (B) Waterfall plot of vertical cross-sections from V$_{\rm{g}}$ = 0 - 0.9V. The integration region is annotated in panel (A). The counts were summed over the emission width for each excitation energy.}
\label{fig:supp-singlet-gate}
\end{figure*} 
In Fig.~\ref{fig:supp-singlet-gate}, we present the results of monitoring the emission channel of the singlet charged exciton X$^{\rm{s}}_-$ for increasing carrier density in a similar manner to Fig. 2/3 for X$_0$ and X$^{\rm{t}}_-$. As noted in the main text, the behavior is the same regardless of the emission channel. We see again that monitoring a different emission channel does not affect the PLE resonances of the 2$s$ states. Both the X$_0^{\rm{2s}}$ and X$^{\rm{2s}}_-$ resonances appear at the same spectral position and show the same spectral dependence with applied gate voltage as discussed in the main text.

\section{X$_-^{\rm{2s}}$  Vertical Cross-Section $g$-factor}
\begin{figure*}[tb]
\includegraphics[width=\linewidth]{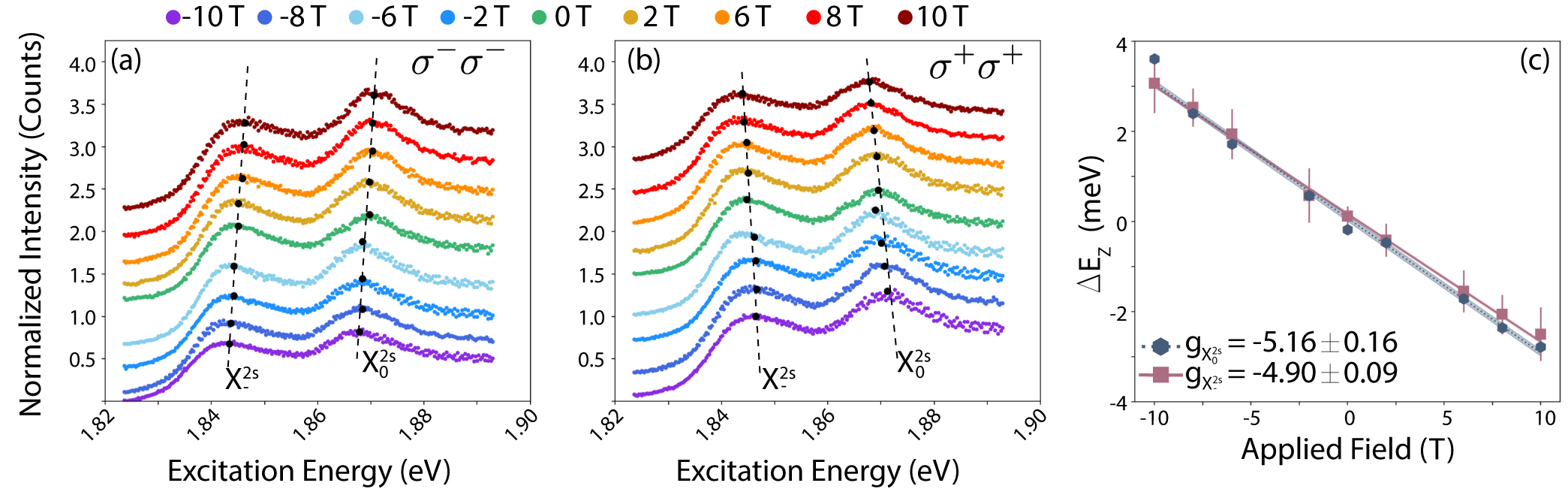}
\caption{Vertical cross-sections in excitation energy as a function of field through the X$_-^{\rm{s}}$ emission channel for (A) ($\sigma^-\sigma^-$) and (B) ($\sigma^+\sigma^+$) marked with the corresponding peak positions for the X$_0^{\rm{2s}}$ and X$_-^{\rm{2s}}$ excitons from fitting.(C) Extracted $g$-factor for X$_0^{\rm{2s}}$ and X$_-^{\rm{2s}}$. The shaded regions in panel (C) on the fit line include the error in the extracted slope.}
\label{fig:fig-singlet-g}
\end{figure*} 

In the main text, we analyzed the magneto-optical dependence of the 2$s$ neutral and charged exciton resonances while monitoring the X$^{\rm{t}}_-$ emission channel. From their valley dependent Zeeman splittings, we extracted corresponding $g$-factors:
\begin{equation}
   g_{\rm{X_0^{2s}}}^{\rm{triplet}} = -5.21\pm0.11 \quad \rm{and} \quad  g_{\rm{X_-^{\rm{2s}}}}^{\rm{triplet}} = -4.98\pm0.11.
\label{eqn:supp-2sXt}
\end{equation}

In this notation, the \emph{triplet} superscript denotes the monitored emission channel while the subscript denotes the exciton resonance for each $g$-factor. 

We expect that changing the monitored emission channel would not change the overall results of the extracted $g$-factors and verify this hypothesis by conducting the same analysis on the X$^{\rm{s}}_-$ emission channel. The results of this analysis are shown in Fig.~\ref{fig:fig-singlet-g}, which is counterpart to Fig. 4 of the main text. We find that $g$-factors extracted from the X$^{\rm{s}}_-$ emission channel for the $2s$ exciton species are: 

\begin{equation}
  g_{\rm{X_0^{2s}}}^{\rm{singlet}} = -5.16\pm0.16 \quad \rm{and} \quad  g_{\rm{X_-^{\rm{2s}}}}^{\rm{singlet}} = -4.90\pm0.09.
\label{eqn:supp-2sXs}
\end{equation}

The $g$-factors extracted from the X$^{\rm{s}}_-$ and X$^{\rm{t}}_-$ emission channels agree within experimental error. This allows us to make the important conclusion that the results from our PLE measurement  are independent of the analyzed emission line.

%
\section{Fitting Procedure for Extracting $g$-factor}\label{sec:g}
\begin{figure*}[tb]
\includegraphics[width=\linewidth]{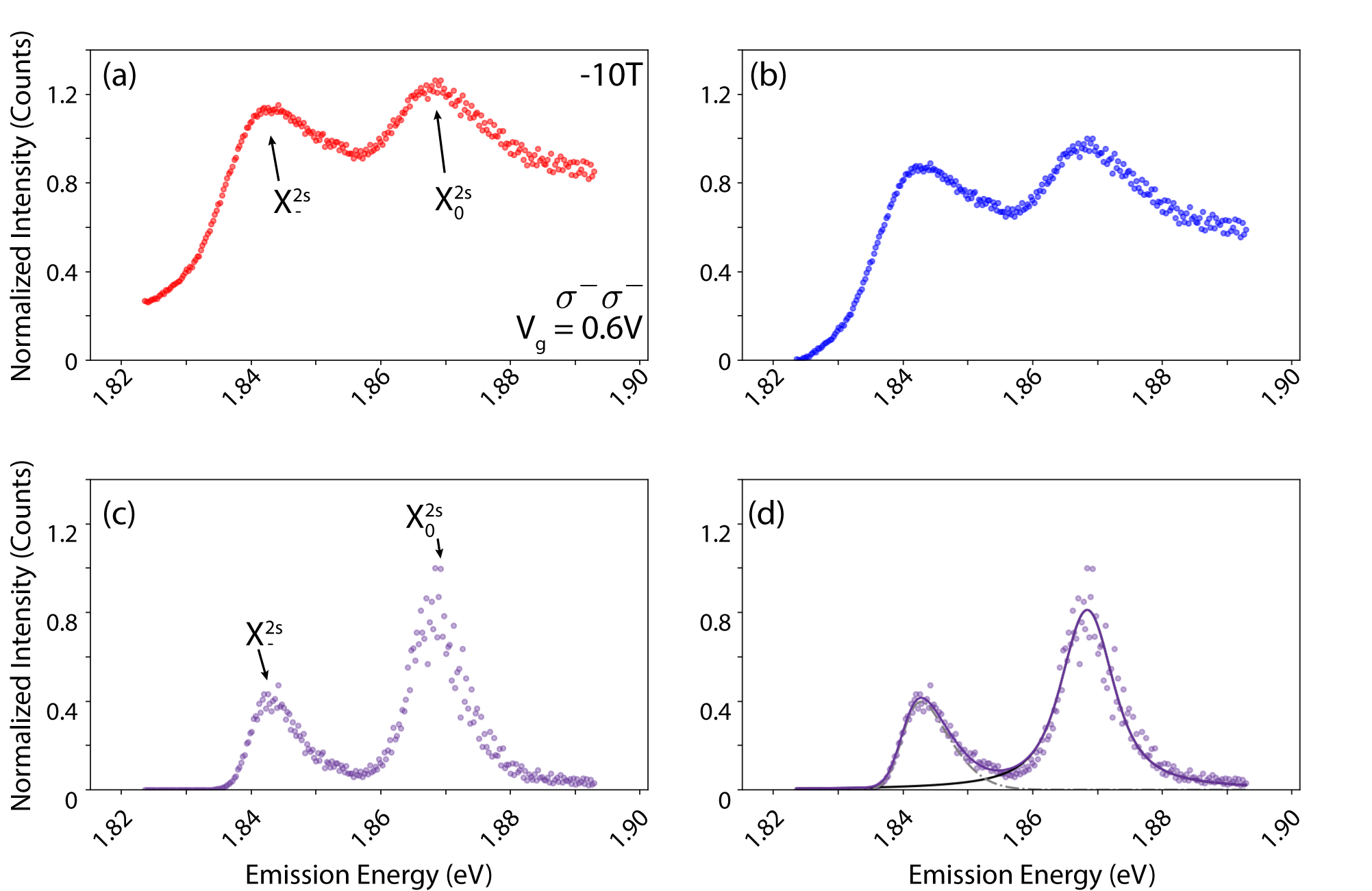}
\caption{(A) Raw Vertical cross-sections of the X$_-^{\rm{t}}$ emission channel at -10T and 0.6 V. The data has been normalized to the maxima of the extracted spectra minus the offset from zero. (B) The same data after the lowest value in the spectra has been subtracted from all points and data has been renormalized to the maxima. (C) Power sharpening of the spectra in panel (B). (D) Peak fitting of the X$_0^{2s}$ and X$_-^{\rm{2s}}$. The denoted input/output polarizations, magnetic field, and gate voltage in panel (A) are the same for all panels.}
\label{fig:fitting-g}
\end{figure*} 
The Zeeman interaction induced splitting is quite small in TMDs, so it is important to take the utmost care to extract accurate peak center information to produce reliable g-factor measurements. In the literature, common techniques for doing this include using the peak maxima or using a weighted-average fitting scheme \cite{Wang-Z}. Both techniques work well for reasonably separated peaks. However, if there is overlap between peak envelopes in a multipeak spectra, both can give erroneous results \cite{Lyons-TP}. In our work, to combat this issue, we introduce use of digital signal processing (DSP) techniques to enhance a peak-fitting approach. Specifically we introduce the ``power law method" \cite{OHaver-T,Dagupta-PK,Wahab-MF}. 

In this technique, the spectra is sharpened by raising each point in the data to a power greater than 1. The result is that peaks ``sharpen" i.e. become narrower and the background/overlap areas between the peaks becomes minimized, effectively giving us a better signal contrast between overlapping peaks. This is namely due to a narrowing of the peaks, so it must be noted that this technique not advisable for extracting information such as full width at half max as a function of a varied parameter. Additionally, it is worth pointing out that while the peaks become narrower the overall shape of the envelope will not change -- i.e. if a peak is symmetric this method will preserve that symmetry just as it would preserve asymmetry in a peak that has an asymmetric envelope. Some work has indicated that excessive peak sharpening in this method can reduce asymmetry but it cannot create or destroy that style of line shape; that is dictated by the underlying physics of the system \cite{Wahab-MF}. Crucially though for our purposes, this method preserves the central signal of the peak which is our desired quantity to extract. To walk through the power-law method used in this manuscript, we turn to Fig.~\ref{fig:fitting-g}.

Fig.~\ref{fig:fitting-g} (A) contains an example of a raw, vertically integrated cut through the X$_-^{\rm{t}}$ emission line with $\sigma^-\sigma^-$ excitation/collection. Since this technique works best with a minimal background, we first subtract the lowest value of the spectra from the entirety of the signal -- Fig.~\ref{fig:fitting-g} (B). Next, we apply a power filter and renormalize the data to the new maxima in Fig.~\ref{fig:fitting-g} (C).

Once the sharpening is performed, the data can be fit with the model of choice. Here, we use a Voigt function to model X$_0^{\rm{2s}}$  and an asymmetric Voigt function to model X$_-^{\rm{2s}}$. While the intrinsic lineshape of an exciton is generally assumed to be Lorentzian, there are many sources of potential inhomogeneous broadening (such as lattice defects, exciton-carrier scattering, or temperature induced thermal broadening) that are generally modelled as a Gaussian envelope. Thus, the Voigt lineshape seems to be the best choice since it is a convolution of these two possible sources of signal and it makes no assumptions about the present sources of broadening \cite{Ajayi-O}. Both in the raw and power-sharpened spectra it is clear that the X$_-^{\rm{2s}}$ state is asymmetric. An asymmetric tail on the lower energy side of a peak is relatively common and is usually indicative of a phonon side band. Higher energy asymmetry is less commonly observed, and generally is attributed to inhomogeneity in the dielectric environment, usually from contamination during the encapsulation process \cite{Lin-Y,Zhu-X}. However, in that case, since optical measurements are local in nature one would expect to see this blue tail on all peaks in a spectra collected at a given location on the sample. Since it is obvious in Fig.~\ref{fig:fitting-g} (C)/(D) that the asymmetry is rather limited to the X$_-^{\rm{2s}}$ signal, we attribute this asymmetry to the presence of two broad, closely spaced peaks that cannot be spectrally resolved. This picture is consistent with the expected presence of a negatively charged doublet as in the 1$s$ state, combined with the reduced intervalley exchange splitting for the 2$s$ charged excitons discussed in the main text as well as in the literature \cite{Arora-A}. 

The fitting was performed using the LMfit library in python, which includes Voigt and asymmetric (skewed) Voigt models as built-in functions \cite{LM-fit}. The library builds on the Levenberg–Marquardt algorithm -- sometimes also referred to as a damped least-squares optimization approach -- for optimizing the fit of a input function to the data provided \cite{Levenberg-K,Marquardt-D}. The repeated fitting during the optimization process by the algorithm allows a standard error to be extracted for all input parameters. Error bars used in the main and supplementary texts for extracted points correspond to the standard error of that point. An example of the  resulting fit using this method is shown in Fig.~\ref{fig:fitting-g} (D).  

%

\section{1$s$ PL and $g$-factor}
\begin{figure*}[tb] 
\includegraphics[width=\linewidth]{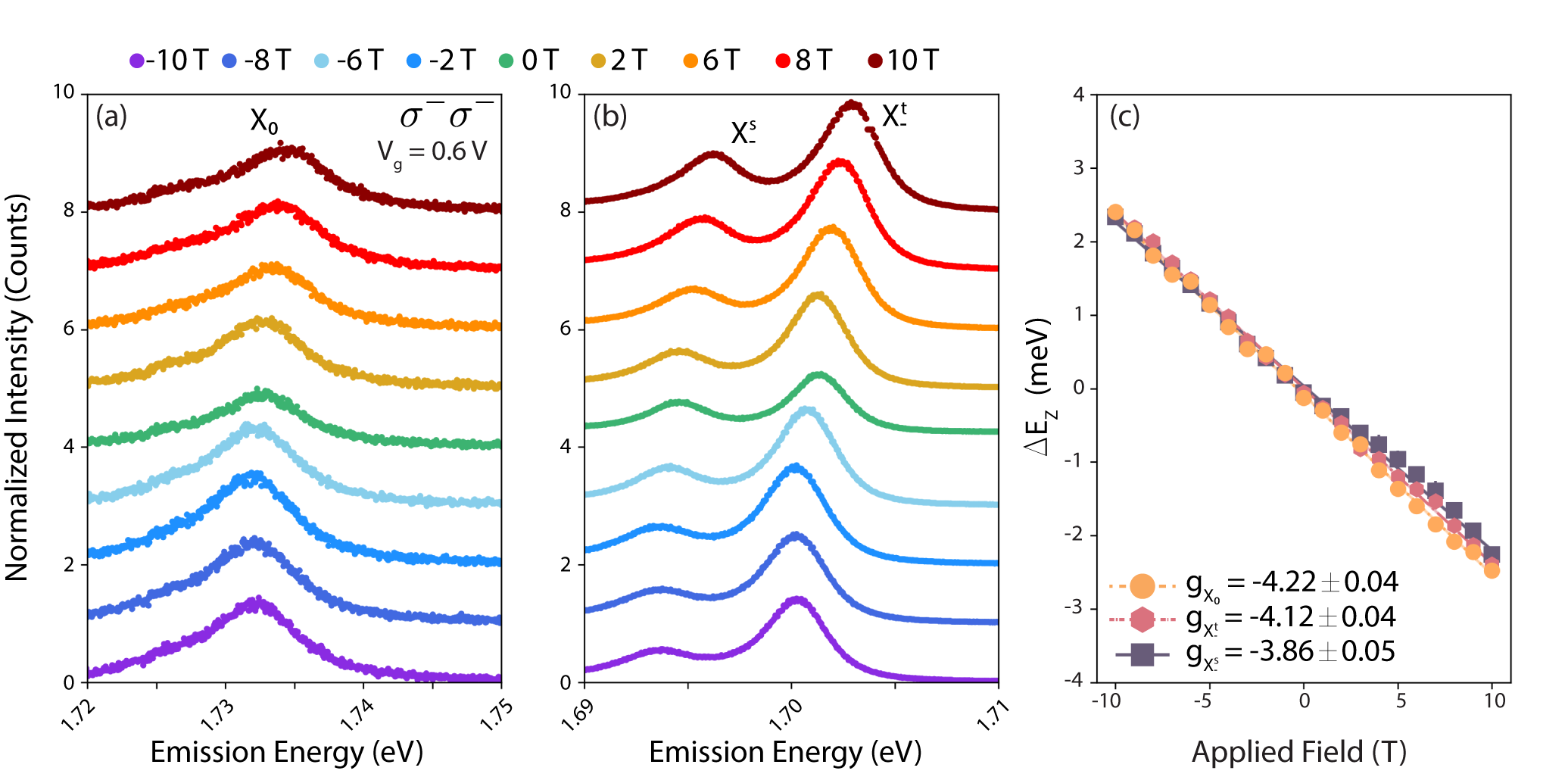}
\caption{1$s$ PL spectra at V$_{\rm{g}}$=0.6 V for the (A) X$_0$, (B) X$_-^{\rm{t}}$/X$_-^{\rm{s}}$. (C) Extracted $g$-factors from fitting PL as a function of field for X$_0$, X$_-^{\rm{t}}$, and X$_-^{\rm{s}}$. For all measurements E$_{\rm{ex}}$=1.92 eV}
\label{fig:fig-1s-g}
\end{figure*}

The vast majority of $g$-factor measurements reported for TMDs in the literature are for 1$s$ states with a limited number of results for X$_0$ Rydberg states. For the purpose of comparison to both the literature and between differing $n$ states, we measured the $g$-factor of X$_0$, X$_-^{\rm{t}}$, and X$_-^{\rm{s}}$ in our system at V$_{\rm{g}}$=0.6 V. This is same carrier environment as our main text measurements of the 2$s$. 

The supplemental measurements of these states were performed using PL with E$_{\rm{ex}}$=1.92 eV. The extracted plots for X$_0$ and X$_-^{\rm{s}}$/X$_-^{\rm{t}}$ are shown in Fig.~\ref{fig:fig-1s-g}(A) and (B), respectively. We note that the relative intensity of X$_-^{\rm{t}}$, and X$_-^{\rm{s}}$ changes from -10T to 10T indicating that there is a degree of valley polarization that is induced by the Zeeman splitting. However, as neither state is fully suppressed, the system is never fully valley polarized.

To extract the peak centers, we use a similar peak-fitting technique as described in \ref{sec:g}, but with the exception of using symmetric Voigt functions to fit each relevant peak. Using this method we find that the $g$-factors of these states are all $\approx -4$:

\begin{equation}
   g_{\rm{X_0}} = -4.22\pm0.04\rm{,} \quad g_{\rm{X_-^{\rm{t}}}} = -4.12\pm0.04\rm{,} \quad \rm{and} \quad g_{\rm{X_-^{\rm{s}}}} = -3.86\pm0.05.
\label{eqn:supp-1sg}
\end{equation}

These results are plotted in Fig.~\ref{fig:fig-1s-g}(C). Our findings are consistent with the relevant literature on WSe$_2$ \cite{Srivastava-A,Li-Z4}. 


%
\section{Estimating the Intercellular Contribution to $g$ for Charged Excitons in Single-Particle Formalism}

\begin{figure*}[tb]
\includegraphics[width=\linewidth]{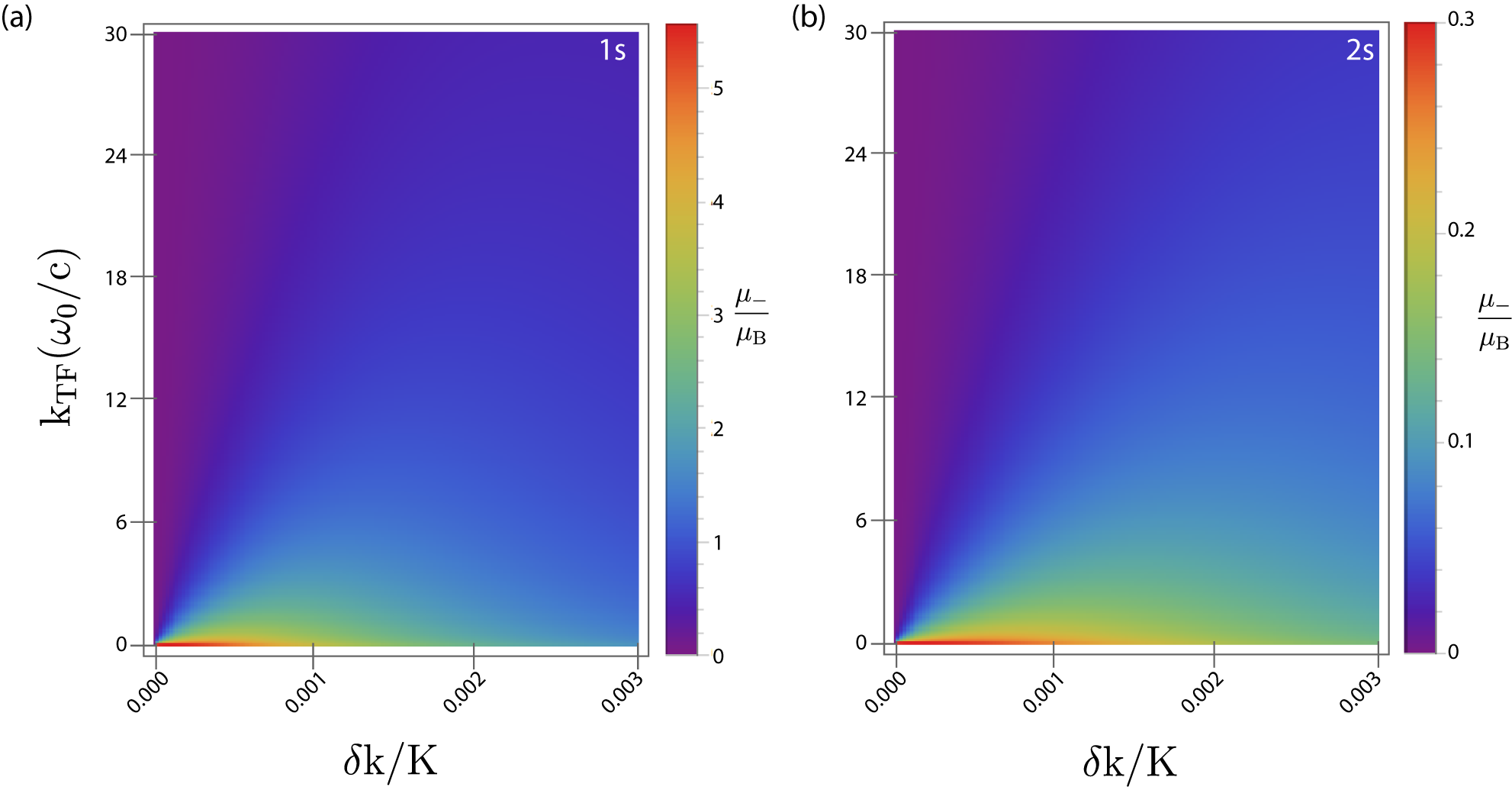}
{\caption{Valley orbital magnetic moment for a charged exciton as a function of the center of mass $k$ from the valley center K for different carrier concentrations (expressed in k$_{\rm{TF}}$) for (A) 1$s$ charged exciton and (B) 2$s$ charged exciton. Note the difference in scale for the resulting magnetic moment in each panel. 
}\label{fig:supp:OMM}}
\end{figure*} 

It is common practice in the literature to use the orbital magnetic moment (OMM) to calculate the intercellular contribution to $g$ for excitons arising from the Berry curvature \cite{Srivastava-A,Chen-SY2,Deilmann-T,Forste-J}. Charged excitons also experience a contribution from the Berry curvature that is induced by the exchange gap near the $\pm$K points \cite{Yu-H}. The resulting OMM is written as,

\begin{equation}
\begin{split}
L_n(k) &= \frac{m}{\hbar}E_{\rm{g}}(k)\Omega_{\rm{X_-}}(k)\\ &=\underbrace{\delta_{\rm{ex}}\bigg(1 + \frac{4J^2k^4}{K^2\delta_{\rm{ex}}^2(k+k_{\rm{TF}})^2}\bigg)^{1/2}}_{E_{\rm{g}}(k)}\underbrace{\frac{2J^2}{{K^2}\delta_{\rm{ex}}^2}\frac{k^2(k+2k_{\rm{TF}})}{(k+k_{\rm{TF}})^3}\bigg(1 + \frac{4J^2k^4}{K^2\delta_{\rm{ex}}^2(k+k_{\rm{TF}})^2}\bigg)^{-3/2}}_{\Omega_{\rm{X_-}}(k)}
\end{split}
\end{equation}

Thus, the charged exciton valley magnetic moment resulting from this exchange interaction is written as,
\begin{equation}
\mu_{\rm{X_-}}(k) = \frac{e}{2m}L(k) = \frac{eJ^2}{\hbar{K^2}\delta_{\rm{ex}}}\frac{k^2(k+2k_{\rm{TF}})}{(k+k_{\rm{TF}})^3}\bigg(1 + \frac{4J^2k^4}{K^2\delta_{\rm{ex}}^2(k+k_{\rm{TF}})^2}\bigg)^{-1}
\label{eqn:OMM}
\end{equation}

 Here, $J$ is the e$^-$/h$^+$ exchange coupling strength, $k_{\rm{TF}}$ is the Thomas-Fermi wavevector for the carrier screening, $\delta_{\rm{ex}}$ is the exchange-induced gap discussed previously, and $K$ is the position of the valleys in $k$-space \cite{Yu-H}.

Fig.~\ref{fig:supp:OMM} shows the results of evaluating Eqn. {\ref{eqn:OMM}} to extract the magnetic moment as a function of the center of mass (COM) momentum $k$ with respect to its distance from the valley center $K$ and Thomas-Fermi wave vector. The Thomas-Fermi wave vector is used as a proxy in the system the carrier concentration present with 10 $\omega_0/c \simeq 10^{11}\rm{cm}^{-2}$ \cite{Srivastava-A,Yu-H}. Panel (A) shows the results for a 1$s$ charged exciton while panel (B) shows the results for a 2$s$ charged exciton. Though the plots look very similar, note the difference in scale for $\mu_{\rm{X_-}}$: the resulting valley OMM for the 2$s$ charged exciton is much smaller than for the 1$s$ charged exciton. This difference largely results from an increase in the Bohr radius, while both $J$ and $\delta_{\rm{ex}}$ decrease.  

\section{Raman Phonon Lines in X$_0$ PLE Emission Response}

Previous experiments exploring the X$^{\rm{2s}}_0$- X$^{\rm{3s}}_0$ energy regime via PLE have demonstrated two prominent Raman modes resulting from electron-phonon coupling between WS$\mathrm e_2$ and hBN that become bright when their emission energies match $\mathrm X_0$ \cite{Chow-C, Serrano-J, Jin-C}; both are labeled in Fig.~1(D) and 2(A) of the main text. The first is an optical phonon, ZO(hBN), that is silent in pure hBN and becomes prominent only when in close proximity to WS$\mathrm e_2$ \cite{Chow-C, Serrano-J, Jin-C}. The second is a combined mode of ZO(hBN) and an out-of-plane optical vibrational mode in WS$\mathrm e_2$, $\rm{A_{1g}}$(WS$\mathrm e_2$) \cite{Jin-C}. We measure the spectral displacement of these two phonon line to be 31.1 meV, which is in good agreement with prior reports \cite{Jin-Z}. 

Since the ZO(hBN) + $\rm{A_{1g}}$(WSe$_2$) phonon has the same energy as the gap between $\mathrm X_0^{2\rm{s}}$ and $\mathrm X_0$, when the laser is tuned to the energy of $\mathrm X_0^{2\rm{s}}$, the Raman signal overlaps the X$_0^{2\rm{s}}$ resonance and becomes orders of magnitude brighter than any other signal from the sample. This degeneracy can be broken through gating as the $\mathrm X_0^{2\rm{s}}$ resonance blue-shifts with increasing carrier density and the Raman modes are unaffected. Fig.~2(A) depicts this shift, showing the spectral isolation of $\mathrm X_0^{2\rm{s}}$ from the ZO(hBN)+$\rm{A_{1g}}$(WS$\mathrm e_2$).